%% file: main.tex
\documentclass[journal]{IEEEtai}

\usepackage[colorlinks,urlcolor=blue,linkcolor=blue,citecolor=blue]{hyperref}

\usepackage{color,array}

\usepackage{graphicx}

\usepackage{amsmath,amssymb,amsfonts}
\usepackage[ruled]{algorithm2e}
\usepackage[numbers]{natbib}
\usepackage{amssymb}
\usepackage{kotex} 
\usepackage{ctable}
\usepackage{setspace}
\usepackage{amsmath}
\usepackage{booktabs}
\usepackage{multirow}
\usepackage{longtable}
\usepackage{pdflscape} 
\usepackage{setspace}

\usepackage{arydshln} 

\DeclareMathOperator*{\argmin}{arg\,min}
\DeclareMathOperator*{\argmax}{arg\,max}

\newcommand{\mytablespacing}{1}

\begin{document}

\title{Security and Privacy Issues in Deep Learning}

\author{Ho Bae$^{\dag}$,
    Jaehee Jang$^{\dag}$,
    Dahuin Jung,
    Hyemi Jang,
    Heonseok Ha,
    Hyungyu Lee,
    and~Sungroh Yoon\textsuperscript{*},~\IEEEmembership{Senior Member,~IEEE}

    \thanks{$^{*}$Corresponding author: Sungroh Yoon (e-mail: sryoon@snu.ac.kr).}
    \thanks{$^{\dag}$: These authors contributed equally to this work.}

    \thanks{This paragraph of the first footnote will contain the date on which you submitted your paper for review. It will also contain support information, including sponsor and financial support acknowledgment. For example, ``This work was supported in part by the U.S. Department of Commerce under Grant BS123456.'' }
    \thanks{H. Bae was with the Seoul National University, Seoul 08870, Republic of Korea. He is now with Ehwa University, Seoul 03760 Republiic of Korea. (e-mail: hobae@ehwa.ac.kr).}
    \thanks{J. Jang, D. Jung, H. Jang, H. Ha, H. Lee and S.Yoon are with the Department of Electrical and Computer Engineering, Seoul National University, Seoul 08870, Republic of Korea. (email: \{hukla, annajung0625, wkdal9512, heonseok.ha, rucy74, sryoon\}@snu.ac.kr).}
}

\markboth{Journal of IEEE Transactions on Artificial Intelligence, Vol. 00, No. 0, Month 2020}
{Ho \MakeLowercase{\textit{et al.}}: Security and Privacy Issues in Deep Learning}

\maketitle

\input{0-abstract.tex}

\section{Introduction}
\label{sec:introduction}
\input{1-introduction.tex}

\section{Secure AI}
\label{sec:secureai}
\input{2-secureai.tex}

\section{Private AI}
\label{sec:privateai}
\input{3-privateai.tex}

\section{Conclusion}
\label{sec:conclusion}
\input{4-conclusion.tex}

\bibliographystyle{IEEEtran}
\bibliography{spai}

\end{document}

%% file: 0-abstract.tex
\begin{abstract}
    To promote secure and private artificial intelligence (SPAI), we review studies on the model security and data privacy of DNNs.
    Model security allows system to behave as intended without being affected by malicious external influences that can compromise its integrity and efficiency.
    Security attacks can be divided based on when they occur: if an attack occurs during training, it is known as a poisoning attack, and if it occurs during inference (after training) it is termed an evasion attack.
    Poisoning attacks compromise the training process by corrupting the data with malicious examples, while evasion attacks use adversarial examples to disrupt entire classification process.
    Defenses proposed against such attacks include techniques to recognize and remove malicious data, train a model to be insensitive to such data, and mask the model's structure and parameters to render attacks more challenging to implement.
    Furthermore, the privacy of the data involved in model training is also threatened by attacks such as the model-inversion attack, or by dishonest service providers of AI applications.
    To maintain data privacy, several solutions that combine existing data-privacy techniques have been proposed, including differential privacy and modern cryptography techniques.
    In this paper, we describe the notions of some of methods, e.g., homomorphic encryption, and review their advantages and challenges when implemented in deep-learning models.
\end{abstract}

\begin{IEEEImpStatement}
    With advancements in deep learning technologies, AI-based applications have become prevalent in various fields.
    However, existing deep learning models are vulnerable to various security and privacy threats.
    These threats can cause grave consequences in real life: for example, if an autonomous vehicle is compromised, the system could fail to recognize a pedestrian owing to an adversary, which can cause a lethal accident.
    This paper systemically categorizes representative threats that occur in deep-learning and defense methods and presents insights for deep learning developers to develop robust AI applications.
\end{IEEEImpStatement}

\begin{IEEEkeywords}
    Private AI, Secure AI, Machine Learning, Deep Learning, Homomorphic Encryption, Differential Privacy, Adversarial Example, White-box Attack, Black-box Attack
\end{IEEEkeywords}

%% file: 1-introduction.tex
\IEEEPARstart{T}{he} development of deep learning (DL) algorithms has transformed the the approach adopted to address several real-life-data-driven problems, such as managing large amounts of patient data for disease prediction~\citep{shickel2017deep}, preforming autonomous security audits from system logs~\citep{buczak2016survey}, and developing self-driving cars using visual object detection~\citep{ren2015faster}.
However, the vulnerabilities of DL-based systems wht respect to security and privacy have been extensively studied to prevent cyberattacks.

If the input data is compromised, a DL-based system can produce in accurate or undesired results.
For example, jamming the sensors ~\citep{yan2016can} or occluding the camera lens~\citep{li2019adversarial} of an autonomous driving system can have dangerous effects on its performance.
Similarly, bio-metric authentication systems using face recognition~\citep{carlini2018audio} can be bypassed by adding noise or digitally editing a pair of glasses onto the image of a face~\citep{sharif2016accessorize} to achieve false positive results.

In this study, we divided such attacks into evasion (inference phase) and poisoning (training phase) attacks.
In previous studies of evasion attack, attacks have typically been categorized as white-box or black-box attacks.
Initially, most forms of evasion attacks were white-box attacks---they require prior knowledge of the DL model parameters and structure---that attempt to subvert the learning process or reduce the classification accuracy by injecting adversarial samples using gradient-based techniques~\citep{biggio2013evasion, goodfellow2014explaining}.
Recently, black-box attacks have become more prevalent; they function by exploiting the classification confidence of the target model to produce incorrect classification information.
Poisoning attacks can also be divided into white- and black-box attacks based on the model accessibility.
However, in this paper, we categorize poisoning attacks into three subclasses based on the vulnerability of the target model: performance degradation, targeted poisoning, and backdoor attacks.

The methods proposed to defend DL-based systems against evasion attacks include empirical approaches---gradient masking~\citep{buckman2018thermometer,dhillon2018stochastic,song2017pixeldefend}, increasing robustness~\citep{goodfellow2014explaining, madry2017towards, zhang2019theoretically}, and detection of attacks~\citep{metzen2017detecting, meng2017magnet, hu2019new}--that can be implemented against known attacks and model certification approaches~\citep{katz2017reluplex, wong2018provable, cohen2019certified}.
We employed defense techniques to counter poisoning attacks separately; they mainly focus on detecting anomalous data~\citep{steinhardt2017certified,koh2017understanding,paudice2018detection,paudice2018label,chen2018detecting,tran2018spectral,chen2018detecting} and making the model robust to poisoning attacks by pruning or fine-tuning with reliable clean data~\citep{liu2018fine,gu2017badnets,wangneural}.

Current DL systems additionally face the threat of privacy breach.
Although it has been demonstrated that recovering or identifying some of the training data~\citep{model_inversion_confidence, shokri2017membership} is possible, a privacy breach can occur in other situations as well.
There are considerable risks involved in training a DL model with data owned by multiple parties; for instance, in the case of deploying an application via a third-party cloud system. Various attempts have been made to counter these threats, by applying conventional security techniques, such as homomorphic encryption, secure multiparty computation, or differential privacy, to DL systems.

In this paper, we review recent studies on model security and data privacy that have contributed towards building a secure and private artificial intelligence (SPAI).
To address the need for robust artificial intelligence (AI) systems, we further compile fragmented findings and techniques with the objective of providing insights relevant to future research.

To summarize, we review recent research on privacy and security issues associated with DL in the following domains.

\begin{enumerate}
    \item \textit{Attacks on DL models} :
          The two major types of attacks on DL relating to different phases---evasion and poisoning attacks---evasion attacks involve the inference phase whereas poisoning attacks involve the training phase.
    \item \textit{Defense of DL models} :
          The various defense techniques proposed, which can be categorized into two large groups based on the type of attack---evasion and poisoning; the techniques implemented against evasion attacks can be further categorized into empirical (e.g., gradient masking, robustness, and detection) and certified approaches.
    \item \textit{Privacy attacks on AI systems} : 
        The potential privacy threats to DL-based systems arising from service providers, information silos and users.
    \item \textit{Defense against a privacy breach} : 
        The most recent defense methods based on cryptography, such as homomorphic encryption, secure multiparty computation, and differential privacy.

\end{enumerate}

%% file: 2-secureai.tex
\input{1_tb-secureai.tex}

In this section, we suggest the concept of secure AI---an AI system with security guarantees---to encourage studies in this field. Additionally, we introduce and taxonomize the groups of studies conducted on security attacks and defense mechanisms, as described in Table~\ref{tbl:secureai-overview}.
\subsection{\textbf{Security Attacks on Deep Learning Models}}
\begin{figure}[t]
    \centering\includegraphics[width=1\linewidth]{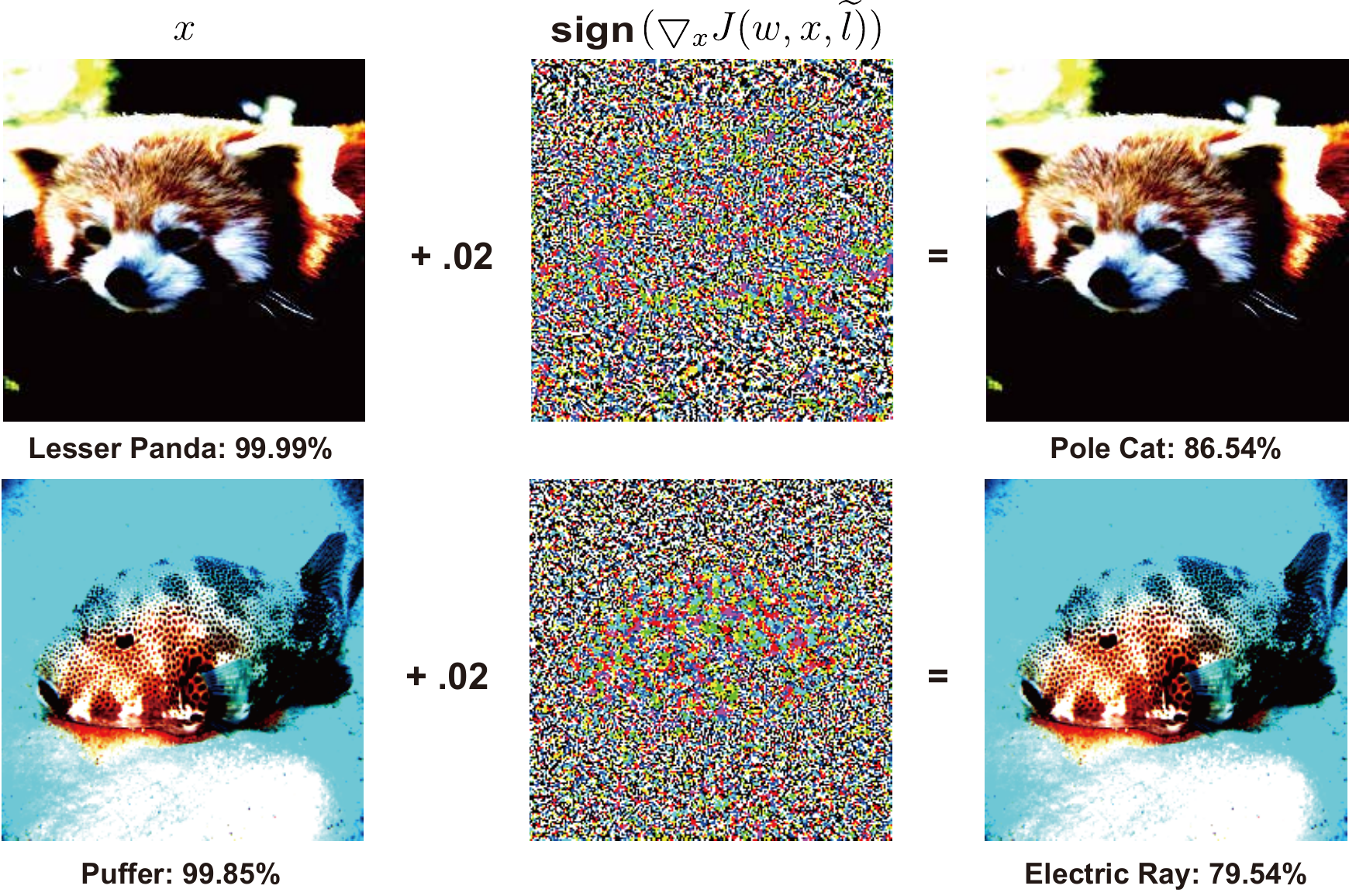}
    \caption{An adversarial example generated by the fast gradient sign method \citep{goodfellow2014explaining}. Left: the original image. Middle: adversarial perturbations. Right: the adversarial image containing the adversarial perturbation.}
    \label{fig:adversarial_attack}
\end{figure}

Table~\ref{tbl:secureai-attack} briefly describes evasion and the poisoning attacks.
As suggested earlier, a poisoning attack attempts to destroy the model during training. The adversarial example used in this attack is known as an adversarial training example, as depicted in Fig.~\ref{fig:adversarial_attack}. In an evasion attack, the adversarial (test) examples are applied during the inference phase, causing the model to misclassify the input. Both types of attacks can be defined as white or black-box attacks.
However, as poisoning attacks are yet understudied, we categorized them as a) performance degradation , b) targeted poisoning , and c) backdoor attacks.

\begin{figure}[t]
    \centering
    \includegraphics[width=0.5\textwidth]{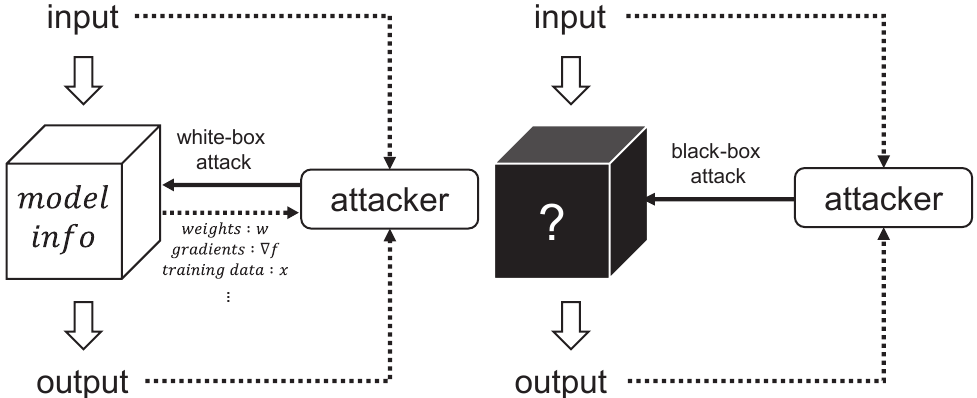}
    \caption{Overview of a white-box (left) and black-box (right) attack.}
    \label{fig:evasion_attack}
\end{figure}

In addition to being classified based on the phase of the workflow that sustains the attack, they vary based on the amount of information available to the attacker.
On one hand, if the attacker gains access to all the information, including the model structure and parameters, their success is highly likely; however, this situation is impractical as demonstrated in Fig.~\ref{fig:evasion_attack} (left). On the other hand, if the adversary has limited information about the model (i.e., no access to ground-truth labels or limited authority), the attacks would be difficult to execute and alternative methods would be required, such as substitute models or data (see Fig.~\ref{fig:evasion_attack} (right)).

There are two approaches to perform an attack: targeted and nontargeted. 
An attack is targeted if its objective is to alter the classifier's output to a specific target label; a nontargeted attack, simply aims to cause inefficient labeling, i.e., to assign a label that holds no value. 
Generally, nontargeted attacks are more successful than targeted attacks.

\subsubsection{\textbf{Evasion Attacks}}

\paragraph{\textbf{White-box attack}}
\label{whitebox_attack}
The first study of evasion attacks~\citep{szegedy2013intriguing} was performed using the \textit{limited-memory Broyden–Fletcher–Goldfarb–Shanno} (L-BFGS) algorithm to generate adversarial examples. Szegedy et al.~\citep{szegedy2013intriguing} proposed a targeted attack method called a box L-BFGS adversary, which involves solving the simple box-constrained optimization problem.
\begin{equation}
    \begin{aligned}
         & \mathbf{minimize}\: \left \| n \right \| _{2}                         \\
         & \:\:\:\:\:\:\: \mathbf{s.t.} \:\:\:\:\:\:\:\:\:   f(x+n) = \tilde{l},
        \label{eq:LBFGS}
    \end{aligned}
\end{equation}
where $f$ is the classifier, $x \in \mathbb{R}^{I^h\times J^w \times K^c}$ is the unperturbed image ($I^h, J^w, \textrm{ and } K^c$ represent the height, width, and the number of channels of the image), $\tilde{l} \in \{1, \cdots, k^c\}$ is the target label, and $n$ represents the minimum amount of noise needed for a model to disassociate an image from its true label. 
The box L-BFGS adversary searches for the minimum perturbation required for a successful attack.
This form of attack has a high misclassification rate and high computational cost because the adversarial examples must be generated by solving Equation \ref{eq:LBFGS}.

Although \textit{Carlini-Wagner’s attack} (CW attack)~\citep{carlini2017towards} is also based on the box L-BFGS attack~\citep{szegedy2013intriguing}, it uses a modified version of Equation~\ref{eq:LBFGS}:
\begin{equation}
    \begin{aligned}
        \mathbf{minimize}\; D\left ( \tilde{x},x \right ) + c\cdot g \left ( \tilde{x} \right ),
    \end{aligned}
\end{equation}
where $\tilde{x}$ is the adversarial example, $D$ is a distance metric that includes $L_{p}$, $L_{0}$, $L_{2}$, and $L_{\infty}$, $g(\tilde{x} )$ is an objective function, in which, $f(\tilde{x} )=\tilde{l}$ if and only if $g(\tilde{x}) \leq 0$, and $c > 0$ is a constant.
Here, the Adam~\citep{kinga2015method} optimizer---adopted to enhance the effectiveness of this attack---conducts a rapid search for adversarial adversarial examples. 
The authors of \citep{carlini2017towards} used the method of changing the variables or the projected gradient descent to support box constraints as a relaxation process after each optimization step. 

Papernot et al.~\citep{papernot2016limitations} introduced a targeted attack method that optimizes within the $L_0$ distance. A \textit{Jacobian-based saliency map attack} (JSMA) is used to construct a saliency map based on a gradient derived from a feedforward propagation, and subsequently modifies the input features that maximize the saliency map such that the probability that an image classified with the target label $\tilde{l}$ increases.

In general, a DL model is described as nonlinear and overfitting; however, the \textit{fast gradient sign method} (FGSM)~\citep{goodfellow2014explaining} is based on the assertion that the main vulnerability of a neural network to adversarial perturbation is its linear nature.
FGSM linearizes the cost function around its initial value, and finds the maximum value of the resultant linearized function following the closed-form equation:
\begin{equation}
    \begin{aligned}
        \tilde{x} = x + \xi \cdot \mathrm{sign}(\nabla_{x}J(w, x, \tilde{l}))
        \label{eq:FGSM}
    \end{aligned}
\end{equation}
where $w$ is the parameter of the model.
The parameter $\xi$ determines the strength of the adversarial perturbation applied to the image, and $J$ is the loss function for training. 
Although this method can generate adversarial examples in a cost-efficient manner, it has a low success rate.
 
Various compromises have been made to overcome the shortcomings of both the above-mentioned attacks. 
One such compromise is the \textit{iterative FGSM} \citep{kurakin2016adversarial}, which invokes the FGSM multiple times, taking a small step after each update, followed by a per-pixel clipping of the image. 
Further, it can be proved that the result of each step will be in the $L_{\infty}$ $\varepsilon$-neighborhood of the original image. 
The update rule can be expressed as follows:
\begin{equation}
    \begin{aligned}
        \tilde{x}_{0} = x,
        \tilde{x}_{N+1} = \mathrm{Clip}_{x, \xi} \left \{ \tilde{x}_{N} + \xi \cdot \mathrm{sign}(\nabla_{x}J(w, \tilde{x}_{N}, \tilde{l})))  \right \},
        \label{eq:I-FGSM}
    \end{aligned}
\end{equation}
where $\tilde{x}_{N}$ is the intermediate result at the N\textsuperscript{th} iteration. 
This method processes new generations more quickly and has a higher success rate.

Noise added to the input data naturally promotes misclassification. 
\textit{Universal adversarial perturbations}~\citep{moosavi2017universal} are image-agnostic perturbation vectors that have a high probability of misclassification with respect to natural images.
Supposing a perturbation vector $n \in \mathbb{R}^{I^h \times J^w \times K^c}$ perturbs the samples in the dataset and that $\mathcal{X}$ represents the dataset containing the samples,
\begin{equation}
    \begin{aligned}
        f\left ( x+n \right ) \neq f\left (x \right )  \text{, for most }\;  x \sim \mathcal{X}.
    \end{aligned}
\end{equation}
The noise $n$ should satisfy  $\left \| n \right \|_{\mathbb{p}} \leq \xi$, and
\begin{equation}
    \begin{aligned}
        \underset{x\sim \mathcal{X} }{\mathbf{P}}\ \left (f\left ( x+n \right ) \neq {f\left ( x \right )}  \right ) \geq 1 - \delta,
    \end{aligned}
\end{equation}
where $f$ is the classifier, $\xi$ restricts the value of the perturbation, and $\delta$ is the fooling rate.

The \textit{backward pass differential approximation} (BPDA)~\citep{athalye2018obfuscated} is an attack that has been claimed to overcome gradient-masking defense methods by performing a backward pass with the identity function to approximate the true gradients of samples.

With the development of adversarial defense methods, more advanced attack methods have been proposed. 
Brendel et al.~\citep{brendel2019accurate} developed a \textit{Brendel and Bethge attack}~(B\&B attack) in which adversarial examples are generated from incorrectly classified regions.
This method uses a combination of a gradient-based attack and boundary attack~\citep{brendel2017decision}, and is a black-box attack. 
It estimates the local boundary between adversarial and clean examples using gradients and moves the adversarial examples close to the clean examples along that boundary. 
Because this method finds optimal adversarial examples by optimization, it can be applied to different adversarial criteria and any norm bound ($L_0, L_1, L_2$, and $L_\infty$).

Recently, Croce et al.~\cite{auto_attack} introduced the auto attack (AA), an ensemble attack that consists of several white-box attacks and one black-box attack~\cite{fab_attack, square_attack}. 
Unlike previous methods, AA uses recursive attacks to separate evaded test data and continues the attack on the remaining data with the successive attack sequentially. 
Consequently, it is more powerful than previous white-box attacks.

\begin{figure*}[t]
    \centering
    \includegraphics[width=0.85\textwidth]{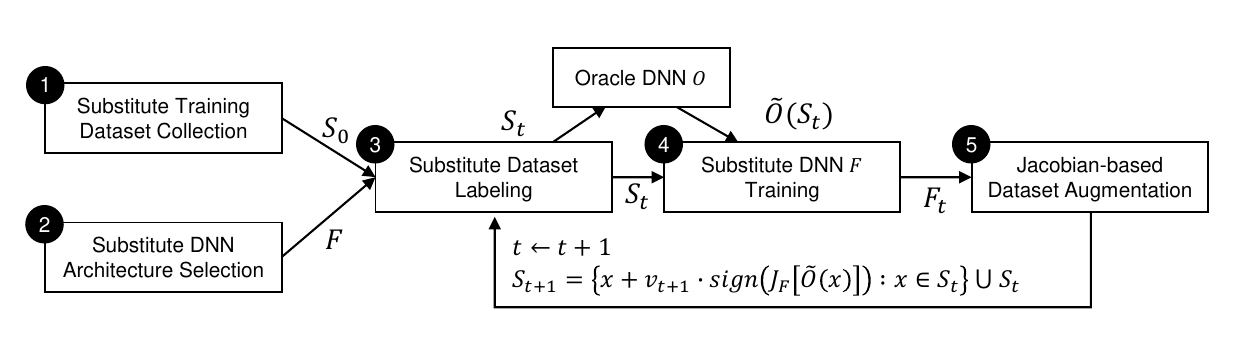}
    \caption{Overview of a practical black-box attack~\citep{papernot2017practical}: the attacker (1) collects a training set $S_0$ for an initial substitute model and (2) selects an appropriate architecture $F$. Using the oracle model $\tilde{O}$, the attacker (3) labels the training set $S_t$ and (4) trains the substitute model $F_t$. Following this, the Jacobian-based adversarial attack algorithm is implemented, the dataset is augmented by the attacker, and steps (3) through (5) are repeated for $t$ epochs.}
    \label{fig:practical_attack}
\end{figure*}

Adversarial examples are typically designed to perturb an existing data point within a small matrix norm at the pixel level, i.e., the samples are \textit{norm-bounded}. 
Most researchers have used this characteristic to propose new defense methods.
Nonetheless, to overcome this drawback, various methods have been proposed that semantically alter the attributes of the input image instead of employing a norm-bounded pixel-level approach.

\textit{Natural generative adversarial network (GAN)}~\citep{zhao2018generating} generates adversarial examples that appear natural to humans. 
Zhao et al. used the latent space $z$ of the GAN structure to search for the required perturbation.
A matching inverter ($\mathrm{MI}$) is used to search for $z^*$, which satisfies the following:
\begin{equation}
    \begin{aligned}
        z^* = \underset{\widetilde{z}}{\text{argmin}} \left \| \widetilde{z} - \mathrm{MI}(x) \right \| \: \text{s.t.} \:  f(G(\widetilde{z})) \neq f(x),
    \end{aligned}
\end{equation}
where $G$ is a generator.
Similarly, Song et al.\citep{song2018constructing} constructed \textit{unrestricted adversarial examples} using an auxiliary classifier GAN (ACGAN)~\citep{odena2017conditional}. Furthermore, they added norm-bounded noise to the generated images to boost the attack ability.

Xiao et al.~\citep{xiao2018spatially} introduced a novel \textit{spatially transformed attack}. 
They used the pixel value and 2D coordinates of each pixel to estimate a per-pixel flow field and generate adversarial examples. 
subsequently, the pixels were moved to adjacent pixel locations along the flow field to produce perceptually realistic adversarial examples. 
They deployed the L-BFGS solver to optimize the following loss function:
\begin{equation}
    \begin{aligned}
         & L_{\mathrm{flow}}(f)=                             \\
         & \sum^{\forall \mathrm{pixels}}_p \sum_{q\in N(p)}
        \sqrt{||\Delta u^{(p)}-\Delta u^{(q)}||^2_2
            +||\Delta v^{(p)}-\Delta v^{(q)}||^2_2
        },
    \end{aligned}
\end{equation}
where $N(p)$ contains the indices of the pixels adjacent to $p$, and $\Delta u^{(\cdot)}$ and $\Delta v^{(\cdot)}$ are the changes in the 2D coordinates of $(\cdot)$. 
The method results in more realistic adversarial examples than those of previous norm-bounded adversarial attacks

Laidlaw et al.~\citep{laidlaw2019functional} used a parameterized function $f$ to generate new pixels for producing adversarial examples.
This method of \textit{functional adversarial attacks} was applied to the color space of images to produce perceptually different but realistic adversarial examples. 
For instance, the method may lighten all the red pixels of an image simultaneously. 
To ensure that adversarial examples are indistinguishable from the original ones, the method minimizes the adversarial loss function and smoothness-constraint loss function similar to previous studies~\citep{carlini2017towards, xiao2018spatially}.

\paragraph{\textbf{Black-box Attack}}
\label{blackbox_attack}
Practically, it is difficult to access the models or training datasets. 
Industrial training models are maintained confidential, and models in mobile devices are not accessible to attackers.
The scenario for a black-box attack is therefore closer to reality: an attacker has no information about the model or the dataset.
However, the input format and the labels outputted by a target model running on a mobile device may be accessible, possibly when the target model is hosted by Amazon or Google.

In a black-box attack, the gradients of the target model are inaccessible to the attackers, who must therefore devise a substitute model.
The attacks performed by means of substitute models are called \textit{transfer attacks}.
It has been shown~\citep{szegedy2013intriguing,goodfellow2014explaining} that neural networks can attack another model without prior knowledge of the number of layers or hidden nodes; however, the task must be known.
This is because a neural network has the linear nature, whereas those of previous studies attribute the transferability to its nonlinearity. 
Activation functions such as sigmoid and ReLU, are known to exhibit nonlinearity. 
The sigmoid function is challenging to implement in learning, whereas ReLU is widely used; however, unlike sigmoid, it does not produce nonlinearity. Thus, a replica of the target model can learn a similar decision boundary for a given task.

The architecture of the substitute model, which may be a convolutional neural network (CNN), a recurrent neural network (RNN), or a multi-layer perceptron (MLP), is approximated based on the input format, which might be images or sequences.
Although the model can be trained by collecting similar data from public sources, the process is highly expensive.

Papernot et al.~\citep{papernot2017practical} addressed this issue by introducing \textit{practical black-box attacks} (Fig.~\ref{fig:practical_attack}), in which an initial synthetic dataset is augmented by a Jacobian-based method. 
This synthetic dataset can be developed from a subset that is not part of the training, data and labeled by inputting it to the target model. 
Thereafter, the trained substitute model can be used to create input data by sending queries to a service such as Google or Amazon; these queries must be severely limited in number and frequency to prevent detection. 
Papernot et al.~\citep{papernot2016transferability} resolved this problem by introducing reservoir sampling, which reduces the amount of data required to train the substitute model.

In addition to being expensive, transfer attacks that employ a substitute model can be blocked by most defense techniques~\citep{tramer2017ensemble}.
Recently, several attacks~\citep{chen2017zoo, brendel2017decision, ilyas2018black, ilyas2018prior, su2019one, guo2019simple} that relied solely on the outputs of the model together with a few queries and other limited information were proposed.

\begin{figure*}[t]
    \centering
    \includegraphics[width=0.85\textwidth]{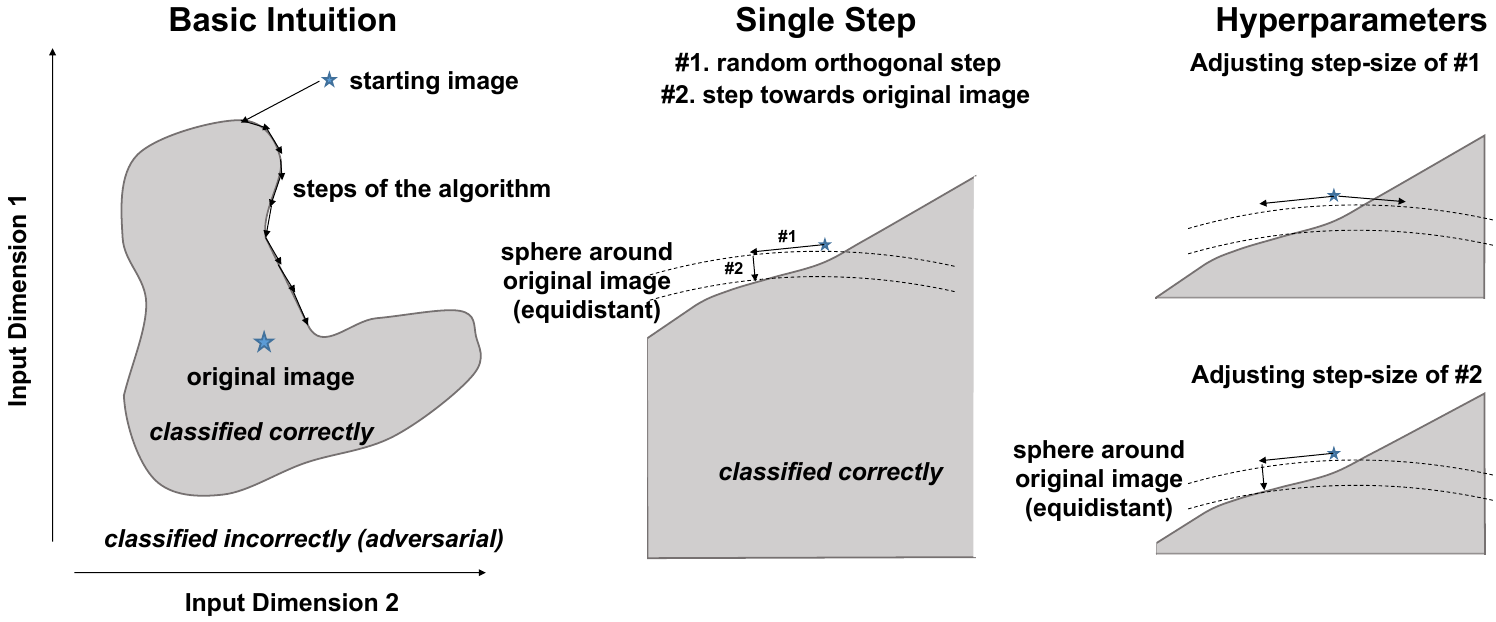}
    \caption{Left: A boundary attack---performs rejection sampling by traversing the boundary between the adversarial and original images. Middle: in each step, the attack determines a new random direction by (\#1) sampling a Gaussian distribution and projecting it on an equidistant sphere, and (\#2) making a small move towards the original image. Right: both two-step sizes are dynamically adjusted to accommodate the boundary~\citep{brendel2017decision}.}
    \label{fig:boundary_attack}
\end{figure*}

Chen et al.~\citep{chen2017zoo} introduced a method for approximating the gradient of a target model that only requires the output score of the target network: they suggested a \textit{zeroth-order optimization} to estimate the gradient of the target model. This method randomly chooses and changes one pixel to compute the adversarial perturbation using zeroth-order optimization with the loss function described in \citep{carlini2017towards}; this process is repeated until sufficient pixels are perturbed. 
The method has been applied successfully to a target network without a gradient, but requires as many queries as the number of pixels. However, attack-space dimension reduction, hierarchical attacks, and importance sampling can be used to reduce the number of queries.

Su et al.~\citep{su2019one} also utilized the score of the network; however, it changed only one pixel in the target image, and is hence called a \textit{one-pixel attack}. 
In this case, a differential evolution algorithm was used to select pixels to perturb. 
These attacks achieved a good success rate. 
Recently, Guo et al.~\citep{guo2019simple} introduced a \textit{simple black-box attack}, which is  query-efficient. 
They developed a method that picks random noise and either adds or subtracts them from an image, the addition or subtraction of random noise was proved to increase the target score of the attack. 
The algorithm repeats this procedure until the attack is successful.

Compared to the methods outlined above, a few other practical attacks rely on predicted labels, because the output scores of the model are usually inaccessible. 
The process of executing a \textit{boundary attack}~\citep{brendel2017decision}---which assumes the worst scenario for attackers---consists of three steps (Fig.~\ref{fig:boundary_attack}): a) an initial sample in the adversarial region is selected; b) a random walk is executed to move the samples toward the decision boundary between the adversarial and non-adversarial regions by reducing the distance to a target example; c) the stages of the walk in the adversarial region are performed by means of rejection sampling. 
Steps b) and c) are then repeated until the adversarial example is sufficiently close to the original image.
Ilyas et al.~\citep{ilyas2018black} introduced a technique similar to a model that requires query-limited, partial-information, and label-only settings. 
Such techniques could implement \textit{natural evolutionary strategies}~(NESs)~\citep{wierstra2008natural} to generate adversarial examples in a query-limited setting. 
Here, an instance of the target class is selected as an initial sample and repeatedly projected onto the $L_{\infty}$-boxes to maximize the probability of the adversarial target class.

\subsubsection{\textbf{Poisoning Attack}}

\begin{figure}[t]
    \centering
    \includegraphics[width=1\linewidth]{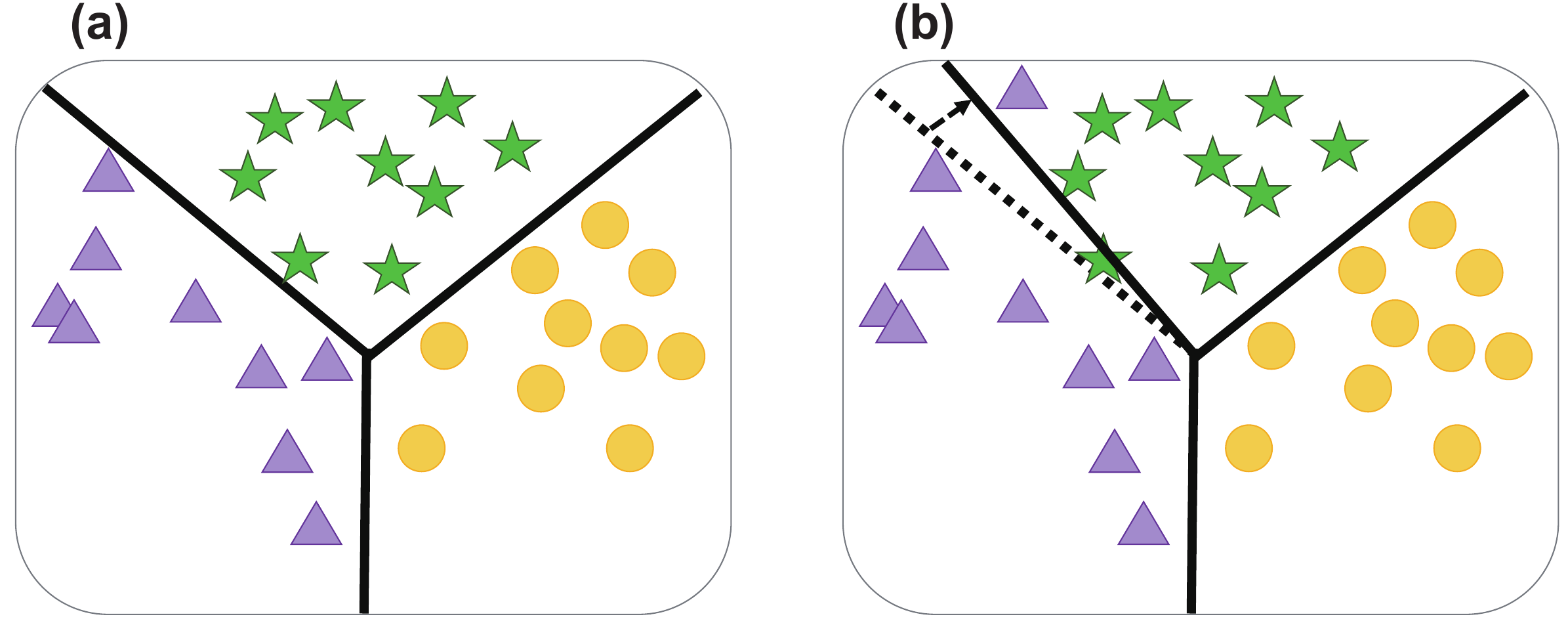}
    \caption{The functionality of poisoning a sample. (a) Decision boundary after training with normal data. (b) Decision boundary after injecting a poisoning sample.}
    \label{fig:poisoning_attack}
\end{figure}

A poisoning attack inserts a malicious example into the training set to interfere with the learning process or facilitate an attack during testing time by changing the decision boundary of the model, as displayed in Fig.~\ref{fig:poisoning_attack}. 
Several poisoning-attack methods applicable to ML techniques, such as SVM or least absolute shrinkage and selection operator (LASSO), can be described mathematically. However, neural networks are difficult to poison owing to their complexity. Nonetheless, the relatively small number of feasible attack methods can be categorized into three types, based on the attacker's goal: performance degradation attacks to compromise the learning process, targeted poisoning attacks to provoke target sample misclassification through feature collision, and backdoor attacks to create a backdoor to be exploited when the system is deployed.

\paragraph{\textbf{Performance degradation attack}}\label{perf_degra_attack}
It aims to subvert the training process by injecting spurious samples generated from a bi-level optimization problem. 
Munoz et al.~\citep{munoz2017towards} described two performance degradation attack scenarios: perfect-knowledge (PK) and limited-knowledge (LK) attacks. 
The PK scenario is an unrealistic setting, and is useful only in a worst-case evaluation. 
In the LK scenario, the attacker typically possesses information, namely $\theta = (\hat{\mathcal{D}},\mathcal{X}, \mathcal{M}, \hat{w})$, where $\mathcal{X}$ is the feature representation, $\mathcal{M}$ is the learning algorithm, $\hat{\mathcal{D}}$ is the surrogate data, and $\hat{w}$ is the learned parameter from $\hat{\mathcal{D}}$, where the $\hat{}$ symbol indicates that the information is partial. 
The bi-level optimization for creating the poisoning samples can be represented as follows:

\begin{equation}
    \begin{aligned}
         & {\mathcal{D}_{c^*} \in \argmax_{\mathcal{D}'_{c} \in \phi(\mathcal{D}_c)}} \qquad
        \mathcal{A}(D'_c,\theta) = J(\hat{\mathcal{D}}_{\mathrm{val}},\hat{w})               \\
         & \mathbf{s.t.} \qquad \hat{w}\in \argmin_{w'\in W}
        J(\hat{\mathcal{D}_{\mathrm{tr}}} \cup \mathcal{D}'_c, w'),
    \end{aligned}
    \label{eq:poison-sa}
\end{equation}
where $\hat{\mathcal{D}}$ is divided into training data $\hat{\mathcal{D}}_{\mathrm{tr}}$ and validation data $\hat{\mathcal{D}}_{val}$. 
The objective function ${A}(\mathcal{D}'_c,\theta)$ evaluates the impact of the poisoning samples among the clean examples.
This function can be considered a loss function, where $J(\hat{\mathcal{D}}_{\mathrm{val}})$ measures the performance of the surrogate model with $\hat{\mathcal{D}_{\mathrm{val}}}$.
The influence of the poisoning sample $\mathcal{D}_c$ is propagated indirectly using $\hat{w}$ following which, the poisoning sample is optimized. 
The primary objective of the optimization is to find a poisoning sample that can degrade the performance of the target model. 
The poison is generic if the target label of the poison sample is arbitrary and not specific.
If a specific target is required, Equation \ref{eq:poison-sa} can be replaced by
\begin{equation}
    \begin{aligned}
        \mathcal{A}(\mathcal{D}'_c,\theta) = -J(\hat{\mathcal{D}}'_{\mathrm{val}},\hat{w}),
    \end{aligned}
    \label{eq:poison-bilevel}
\end{equation}
where $\hat{\mathcal{D}}'_{\mathrm{val}}$ is the manipulated validation set, which is similar to $\hat{\mathcal{D}}_{\mathrm{val}}$, except for the presence of misclassified labels that can produce a desired output.
Munoz et al.~\citep{munoz2017towards} proposed the back-gradient method to solve Equations~\ref{eq:poison-sa} and \ref{eq:poison-bilevel} and generate poisoning examples as an alternative to gradient-based optimization. 
It requires a convex objective function and a Hessian-vector product, which are not produced using complicated learning algorithms, unlike those used to develop neural networks. 
In contrast, Yang et al.~\citep{yang2017generative} were able to apply a gradient-based and GAN-like generative method to deep neural networks (DNNs) using an autoencoder to compute the gradients reduced the computation time by a factor of over 200.

The attacks described above can be detected easily by outlier detection. 
Nonetheless, Munoz et al.~\citep{munoz2019poisoning} recently proposed a GAN-based attack designed to avoid detection. 
Their pGAN model \citep{munoz2019poisoning} had a generator, discriminator and target classifier. A min--max game between the generator and discriminator generated spurious yet realistic images with a poisoning ability. 
A hyperparameter that can adjust the realism and poisoning ability of spurious images affected the trade-off between effectiveness and detectability. 
When the influence of the realistic image-generation process was high, the attack success rate was low. 
Conversely, when the influence of the poisoning ability was high, the generator tended to produce outliers; hence, attacks were more detectable.

\paragraph{\textbf{Targeted poisoning attack}}\label{targted_poisoning_attack} 
This method was introduced by Koh et al.~\citep{koh2017understanding} to cause target test samples selected from the test dataset to be misclassified during the inference phase.
The complexity of neural networks renders identifying the source for classification and explaining the classification in terms of training data challenging. 
Because of the expense of retraining a model after modifying or removing a training sample, the authors formulated the influence of up-weighting or modifying a training sample during training in terms of changes to the parameters and loss functions. 
The attack was optimized based on the amount of change in the test loss caused by the change in the training sample.

Although only a small number of attacks are performed on the training data, the attack may be unsuccessful if the training data are impeded by domain experts. 
Shafahi et al.~\citep{shafahi2018poison} introduced a clean-label attack to circumvent this problem. 
Here, the feature-conflict method was used to ensure that the labels introduced in the attack were appropriate for the images to which they were attached. 
Subsequently, the attacker would select the target image $t$ and base image $b$ from the test set where the target image would be expected to be misclassified as the label of the base image. 
The attack $p$ is initialized with the base image and created using the following equation:

\begin{equation}
    \begin{aligned}
        p=\argmin_{x} ||f(x)-f(t)||^2_2+\beta||x-b||^2_2.
    \end{aligned}
\end{equation}
The attack is generated by optimizing a sample similar to the base image in the image space and close to $ t $ in the feature space mapped by function $f$. The attack surrounds the target feature $f(t)$ and changes the decision boundary, to have the target image categorized within the base class.
For example, if $b$ is a picture of a dog and $t$ is a picture of a bird, the attack changes the decision boundary by adding $p$, a perturbed version of $b$, to the training data. 
As a result, $t$ is erroneously put into class $b$, and $t$ can be used to deceive the classifier.

Shafahi et al.~\citep{shafahi2018poison} analyzed attacks in two of the retraining situations: end-to-end learning which fine-tunes the entire model, and transfer learning, which fine-tunes the final layer. 
They used a one-shot kill attack that generates a poisoning attack from pair of a base and target image through the feature collision method.
The one-shot kill attack was successfully applied to transfer learning after significant changes were made to the decision boundary; however, it was not applied to end-to-end learning, which retrains the lower layers that extract fundamental features.
Shafahi et al.~\citep{shafahi2018poison} succeeded in an attack on end-to-end learning using a watermarking method in which the target image was projected onto a base image by adjusting its opacity and using several target and base images.

Because all neural networks do not have the same feature mapping function, the feature collision using a model cannot be applied to an unknown neural network. 
Zhu et al.~\citep{zhu2019transferable} proposed a feature collision attack~(FC attack) using an ensemble model and a convex polytope attack~(CP attack). 
FC attack adopts the same mechanism of that of \citep{shafahi2018poison}, except for the number of models for feature collisions, which is greater. 
The FC attack was unsuccessful because the constraints on the attack increase, and the attack simply approaches the target in the feature space without changing the predicted result of the target image. 
In contrast, the CP attack using convex properties efficiently transforms the target into or near the polytope. 
However, the attacker would face difficulty poisoning the unknown target model if the model learned new feature mapping functions through end-to-end training. 
Thus, these authors also proposed a multilayer convex polytope attack that generated poisoning attacks using feature collisions of every activation layer.
Moreover, recently, MetaPoison~\citep{huang2020metapoison} generated clean-label data poisoning, which works in an end-to-end setting with bi-level optimization. 
Geiping and Jonas. et al.~\citep{geiping2020witches} succeeded in training their model from scratch on a full-sized, poisoned imageNet dataset using gradient matching.

\paragraph{\textbf{Backdoor attack}}\label{backdoor_attack} 
It is an attack that aims to install a backdoor to be accessed at classification time and was introduced by Gu et al.~\citep{gu2017badnets}, who inserted patches into an image to cause false classifications, such as replacing a stop sign with a speed limit. 
Trojaning attacks \citep{liu2017trojaning} rely on the fact that neural network developers often download pre-trained weights from ImageNet for training or outsource the entire amount of data to suppliers of machine learning as a service (MLaaS). 
In a worst-case scenario, an attacker can directly change the user's model parameters and training data; however, they cannot access the validation set of the user or use the training data to launch attacks. 
Trojaning attacks insert a trigger, in the form of a patch or watermark into an image, which causes it to be categorized into the target class. 
They involve four steps: 1) the trigger and the target class are selected; 2) the attacker selects the node in the target layer with the highest connectivity from the preceding layer of the trained model, and the trigger is updated from the gradient derived from the difference in the activation result of the selected node and the targeted value of the node (the target value is set by the attacker to increase the relatedness of the trigger and the selected node of the target layer); 3) using the mean image of a public dataset, the training data are reverse-engineered to ensure that the images would be classified into the target class; 4) the target model is trained using the reverse-engineered image dataset, including the one containing the trigger. 
When the retrained model is deployed, the image with the trigger is misclassified with the target label. 
Such attacks have been successfully applied to face recognition, speech recognition, auto-driving, and age-recognition applications.

Chen et al.~\citep{chen2017targeted} introduced two strategies to obtain access to a face recognition system, under three constraints, namely 1) no knowledge of the model, 2) access to only a limited amount of training data, and 3) poisoning data not visually detectable. 
In the input-instance-key strategy, a key image is prepared and associated with a target label. 
To model the camera effects, noise is added to the key image. 
The second, pattern-key strategy, has three variants. 
The first strategy, blended injection, combines a spurious image or random pattern within the key image; the output images of this process usually appeared unrealistic. 
The second strategy, accessory injection, applies an accessory, such as glasses or sunglasses, to the key image. 
This is a simple attack to execute during the inference stage. 
The third method, blended accessory injection, combines the first and second strategies. 
Unlike previous studies, in which poisoning data accounted for 20\% of the training data, the authors of \citep{chen2017targeted} only added five poisoned images to 600,000 training images in the input-instance-key strategy, and approximately 50 for the pattern-key strategy. In both cases a backdoor was successfully created.

However, attacks wth visible triggers can be detected easily by human inspection; therefore, recent attacks on image classification introduced invisible triggers such as scattered triggers~\citep{li2019invisible}, which are distributed across the image, warping-based triggers~\citep{nguyenwanet}, and reflection triggers~\citep{liu2020reflection}.
Backdoor attacks performed without label poisoning have also been proposed to increase the stealthiness of a target model~\citep{barni2019new, turner2019label}.
Furthermore, a backdoor can be established by flipping only several vulnerable bits of weights~\citep{liu2017fault,rakin2020t,rakin2020tbt,zhao2019fault}.

\subsection{\textbf{Defense Techniques against Deep Learning Models}}
Defense techniques against both poisoning and evasion attacks have been developed, and the latter can be further categorized into empirical defenses against known evasion attacks and certified defenses, which are also probably effective.

\subsubsection{\textbf{Defense techniques against evasion attacks}}
Various methods have been proposed to defend DL-based systems against evasion attacks (adversarial attacks). 
For example, Kurakin et al.~\citep{kurakin2016adversarial2} suggested that adversarial training can be employed when security against adversarial examples is a concern, which increases robustness against evasion attacks. 
By including adversarial training, defense techniques can be broadly divided into three categories: gradient masking, robustness, and detection.

\paragraph{\textbf{Gradient masking}}
\label{gradient_masking}
Gradient masking obfuscates the gradients used in attacks~\citep{athalye2018obfuscated}. 
There are three approaches predominantly adopted in this method: shattered gradients, stochastic gradients, and vanishing/exploding gradients.

Neural networks generally behave in a largely linear manner \citep{athalye2017synthesizing}. 
As image data is multidimensional in nature, the property of linearity can have adverse effects on the classification, rendering the model vulnerable to adversarial attacks. 
The \textit{shattered gradients} approach involves making the model nondifferentiable or numerically unstable, to ensure that accurate gradients cannot be obtained.
One version of the shattered gradient defense involves \textit{thermometer encoding} \citep{buckman2018thermometer}.
This method applies nondifferentiable and non-linear transformations to the input by replacing one-hot encoding with thermometer encoding.
The thermometer $\tau(j) \in \mathcal{R}^{K}$, can be expressed as follows:

\begin{equation}
    \begin{aligned}
        \tau(j)_l =
        \begin{cases}
            1, & \textrm{if  }\quad l \geq j \\
            0, & \textrm{otherwise}
        \end{cases} .
        \label{eq:thermometer}
    \end{aligned}
\end{equation}
Subsequently, the thermometer discretization function $f$ for a pixel $i \in \{i, \cdots, n\}$ can be defined as
\begin{equation}
    \begin{aligned}
        f_{\mathrm{therm}(x)_i} = \tau(b(x_i)) = \mathcal{C}(f_{\mathrm{onehot}}(x_i)),
        \label{eq:thermometer2}
    \end{aligned}
\end{equation}
where $\mathcal{R}$ is the cumulative sum $C(c)_l = \sum_{j=0}^{l}c_l$, and $b$ is the quantization function. 
Other defense techniques based on gradient shattering include local intrinsic dimensionality (LID)~\citep{ma2018characterizing} metrics or input transformations~\citep{guo2018countering} such as image cropping, rescaling~\citep{graese2016assessing}, bit-depth reduction~\citep{xu2017feature}, JPEG compression~\citep{kinga2015method}, and total variance minimization~\citep{rudin1992nonlinear}.

The \textit{stochastic gradients} approach obfuscates gradients in the inference phase by dropping random neurons in each layer. 
The network then stochastically prunes a subset of the activations in each layer during the forward pass.
Stochastic activation pruning~\citep{dhillon2018stochastic} is a variant of this method in which, the dropout follows the probability from a weighted (rather than uniform) distribution. 
The surviving activations are scaled up to normalize the dynamic range of the inputs to the subsequent layer. 
The probability of sampling the $j$th activation in the $i$th layer is given by
\begin{equation}
    {p^i}_j = \frac{|(h^i)_j|}{\sum^{a^i}_{k=1}|(h^i)_k|},
\end{equation}
where $h^i \in \mathbb{R}^{a^i}$ and $(h^i)_j$ are the values of the $j$th activation in the $i$th layer.
Xie et al.~\citep{xie2017mitigating} also used a randomization technique that inserts a layer in front of the input to the neural network, which rescales and zero-pads the input.

The \textit{vanishing/exploding gradients} method renders the model unusable by deep computation, which restores adversarially perturbed images to clean images. 
These images are then fed to the unmodified classifier. 
\textit{PixelDefend}~\citep{song2017pixeldefend} is a defense algorithm that uses PixelCNN~\citep{oord2016pixel} to approximate the training distribution. 
PixelCNN is a generative model designed to produce images that track the likelihood over all pixels by factorizing it into a product of conditional distributions:
\begin{equation}
    \begin{aligned}
        \textrm{P}_{\mathrm{CNN}}(x) = \prod_{i}\textrm{P}_{\mathrm{CNN}}(x_i|x_{1:(i-1)}).
        \label{pixel_cnn}
    \end{aligned}
\end{equation}
Defense-GAN~\citep{samangouei2018defense} is a similar method that uses a GAN instead of PixelCNN. 
The trained generator projects images onto the manifold of the GAN, and these projected images are then fed into the classifier.

Gradient-based defense algorithms based on the gradient of the initial version are inherently vulnerable to gradient-based attacks.
Athalye et al.~\citep{athalye2018obfuscated} used projected gradient descent to set a perturbation $\upsilon$, combined with the $l_2$ Lagrangian relaxation approach~\citep{carlini2017adversarial}.
Gradient-masking techniques, which exploit obfuscated gradients, are vulnerable to strong gradient-based attacks \citep{kurakin2016adversarial, madry2017towards, carlini2017adversarial}.
Alternatively, an attacker may simply use a different attack~\citep{carlini2017towards,athalye2018obfuscated} to bypass such a defense or the way circumvented by any adversary who uses the true adversarial examples~\citep{he2017adversarial,uesato2018adversarial}. 

\begin{figure*}[t]
    \centering
    \includegraphics[width=0.85\textwidth]{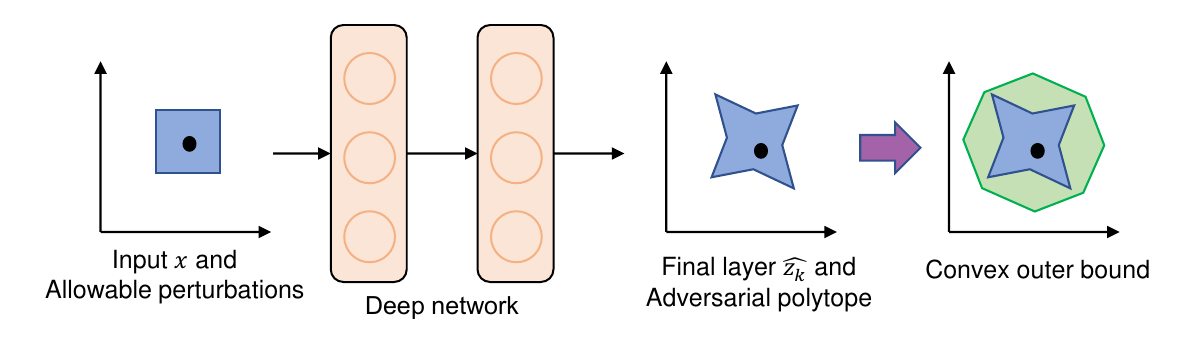}
    \caption{An adversarial polytope and its outer convex bound~\citep{wong2018provable}.}
    \label{fig:adversarial_polytope}
\end{figure*}

\paragraph{\textbf{Robustness}}
\label{robustness}
Gradient obfuscation could prove to be useless in a white-box setting, where increasing robustness may be a better approach. 
One method to increase robustness of the model is to enable it to produce similar outputs from clean and adversarial examples and either by penalizing the difference between them or regularizing the model to reduce the attack surface. 

Most studies on robustness involve \textit{adversarial training}~\citep{goodfellow2014explaining}, which can be viewed as minimizing the worst error caused by the perturbed data of an adversary. 
It can also be perceived as learning an adversarial game with a model that requests labels for the input data.
Other techniques include the \textit{distillation training}~\citep{papernot2016distillation} method which provides robustness to saliency map attacks~\citep{papernot2016limitations}, and a layerwise regularization method~\citep{cisse2017parseval}, which controls the global Lipschitz constant of a network. 
However, none of these methods produce fully robust models and could be bypassed by a multi-step attack, such as the projected gradient descent (PGD).

Most of the optimization problems in ML are solved using first-order methods and variants of stochastic gradient descent; thus, a universal attack can be designed using first-order information. 
Madry et al.~\citep{madry2017towards} suggested that local maxima for the worst error can be found by PGD, on the basis that a trained network that is robust against PGD adversaries will also be robust against a wide range of attacks that assume first-order optimization.

Adversarial training was originally used to train a small model using the MNIST dataset~\citep{goodfellow2014explaining}. 
Kurakin et al.~\citep{kurakin2016adversarial2} extended that work to ImageNet~\citep{deng2009imagenet} using a deeper model with a batch normalization step. 
The relative weights of the adversarial examples can be independently controlled in each batch using the following loss function:
\begin{equation}
    \begin{aligned}
        Loss = \frac{1}{(m-k)+ \lambda k} (\sum_{\mathrm{CLEAN}} J(x_i | l_i) + \lambda \sum_{\mathrm{ADV}} J(\tilde{x_i} | l_i) ),
        \label{eq:LOSS}
    \end{aligned}
\end{equation}
where $J(x|l)$ is the loss on a single example $x$ with true class $l$, $m$ is the total number of training examples in the batch, $k$ is the number of adversarial examples in the batch, and $\lambda$ is the weight applied to adversarial examples.

Defense techniques that change the target function by introducing regularizers or modifying the architecture of the model help increase the robustness of the model against adversarial attacks.
Kannan et al.~\citep{kannan2018adversarial} introduced adversarial logit pairing (ALP), which produces regularization by reducing the distance between the logits of clean examples and those of adversarial examples. The loss function of training then becomes:
\begin{equation}
    J(\mathbb{M}, w)+\lambda \frac{1}{m}
    \sum_{i=1}^{m} L\big(f(x^{(i)};w),f(\tilde{x}^{(i)};w)\big),
\end{equation}
where $J(\mathbb{M}, w)$ is the cost of training a minibatch $\mathbb{M}$, $w$ is the model parameter, and $L$ is the distance function. 
The results showed that a simple regularizer can improve the robustness of a model thatch is trained adversarially.
\textit{Double backpropagation}~\citep{ross2018improving}, which is a regularizer that penalizes the magnitudes of the input gradients, reduces the sensitivity of the divergence between the predictions and uniform uncertainty produced by evasive attacks. 
Miyato et al.~\citep{miyato2018virtual} introduced a regularizer that reduced the Kullback--Leibler divergence between clean and adversarial examples, to render the distributions of the resulting outputs more similar to each other. 
Xie et al.~\citep{xie2019feature} denoise feature maps by adding blocks, such as non-local mean blocks, to a network to reduce adversarial perturbations from the inputs. 
A more recent regularizer~\citep{qin2019adversarial} makes a model behave linearly in the vicinity of the input data, which reduces the effect of gradient obfuscation and improves robustness to adversarial examples.

There are several variants of adversarial training, such as the augmentation of training data or introduction of loss functions. 
Tramer et al.~\citep{tramer2017ensemble} proposed \textit{ensemble adversarial training} to defend against black-box attacks by using adversarial examples generated by other networks. 
Decoupling adversarial example generation from the trained model increases the diversity of the training data. 
In another study, \textit{tradeoff-inspired adversarial defense via surrogate-loss minimization}~ (TRADES)~\citep{zhang2019theoretically} identified a trade-off between adversarial robustness and accuracy. 
The expected errors in adversarial examples are decomposed into the sum of the expected errors in clean examples and a boundary error that corresponds to the likelihood of the closeness of the inupt features to the perturbation-extension of the decision boundary. 
Both these errors are expressed by a surrogate loss function such as cross-entropy or 0-1 loss functions, to yield the following minimization:
\begin{equation}
    \min_{f}\mathbb{E}\{\phi(f(\boldsymbol{x})l)+\max_{\boldsymbol{x'}\in \mathbb{B}(\boldsymbol{x,\xi})}
    \phi(f(\boldsymbol{x})f(\boldsymbol{x}')l / \lambda) \},
\end{equation}
where $\phi$ is the surrogate loss function that represents the expected errors, and $\mathbb{B}(\boldsymbol{x,\xi})$ represents a neighborhood of $x: \{x'\in X : ||x'-x||\leq\xi\}$, which is the expected error and the boundary error weighted by $\lambda$. 
This method showed state-of-the-art performance under both black-box and white-box attacks.
Zhang et al.~\citep{zhang2019defense} proposed a feature scattering-based adversarial training approach that utilized the optimal transport distance between the input data and its adversarial examples for training without label leaking~\citep{kurakin2016adversarial}.
Recently, Zhang et al.~\cite{zhang2020attacks} attempted to to handle the trade-off between standard accuracy and robust accuracy by weakening the adversarial examples during adversarial training. 
They used early-stopped PGD to prevent the adversarial examples used for training from significantly destroying the generalization ability.

According to Schmidt et al.~\citep{schmidt2018adversarially}, the adversarial robustness requires more data to produce successful results.
There are several methods~\cite{carmon2019unlabeled, oat} that use data augmentation methods to increase robustness. 
Carmon et al.,\citep{carmon2019unlabeled} used unlabeled data to improve the robustness; the data were pseudo-labeled by the classifier before being deployed in adversarial training. 
Lee et al.,\cite{oat} proposed out-of-distribution data augmented training (OAT). 
They used out-of-distribution data for training with a uniform distribution label and achieved improved robustness by removing the contributions of undesirable features.

\paragraph{\textbf{Detection}}
\label{detection}
Retaining the ability to detect attacks at the inference phase is considered equally (if not more) valuable to increase the security of a DL-based system and ensure that corrupted input can be rejected.
Most detection methods require no change to the classifier; therefore, they are easy to implement, and can be combined with other defenses.

Metzen et al.~\citep{metzen2017detecting} detected adversarial examples using a binary detector network, which was trained to classify inputs into clean and perturbed examples. 
Using a similar scheme, Meng et al.~\citep{meng2017magnet} separated a detector and reformer network, which were then used to reconstruct clean input.
These networks identified adversarial examples from the reconstruction error, which yields the Jensen--Shannon divergence of the original and reconstructed inputs:
\begin{equation}
    JSD(P || Q)=\frac{1}{2}D_{KL}(P||M)+\frac{1}{2}D_{KL}(Q||M),
\end{equation}
where $P$ is the output resulting from the original inputs, $Q$ is the output of the reconstructed input, and $M$ is the mean of $P$ and $Q$, $M=\frac{1}{2}(P+Q)$.

\textit{Feature squeezing}~\citep{xu2017feature} reduces the search space available to attackers by squeezing the inputs before comparing the prediction results obtained from the squeezed examples with those of the clean examples. 
If there are substantial differences, the original input is likely to be adversarial. 
Squeezing was achieved by color-depth reduction and spatial smoothing (both local and non-local smoothing). 
This method was able to detect adversarial examples in various types of evasion attacks with a low false-positive rate.

Grosse et al.~\citep{grosse2017statistical} identified adversarial examples by applying \textit{statistical metrics} to the output of a classifier. 
They also introduced a method of \textit{integrated outlier detection} in which a classifier is trained to categorize adversarial examples into a new class. 
This involves a small reduction in classification accuracy but a high detection rate.
Feinman et al.~\citep{feinman2017detecting} also detected adversarial examples by examining statistical metrics such as the density of the feature space and Bayesian uncertainty estimates.

Pang et al.~\citep{pang2018towards} minimized reverse cross-entropy as the loss function used to train the model, to identify adversarial examples. 
The reverse cross-entropy loss value of an input $x$ over a label $y$ is expressed as follows:
\begin{equation}
    L^R_{CE}(x,l)=-R_l^{\intercal} \log \sigma(x),
\end{equation}

\begin{equation}
    \begin{aligned}
        R_y=P_i^\lambda =
        \begin{cases}
            \frac{1}{\lambda+1},              & \textrm{}i=y      \\
            \frac{\lambda}{(L-1)(\lambda+1)}, & \textrm{}i \neq y
        \end{cases},
    \end{aligned}
\end{equation}
where $\sigma(\cdot)$ is ;he softmax output, $R_y$ represents the reversed form of the label $y$, and $\lambda$ is the hyperparameter with $\lambda=\infty$ in the experiment. 
In a recent study, Hu et al.,~\citep{hu2019new} introduced a detection method with safety criteria: robustness against random noise and susceptibility to adversarial noise, which is represented as robustness against Gaussian noise and the minimum number of steps required to perturb the input, respectively.
They achieved unprecedented accuracy in a white-box setting.

\paragraph{\textbf{Certified approach}}
\label{certification}
The robustness of most defenses can only be established empirically in the context of known types of attacks. 
An empirically robust classifier may be overcome by new and stronger attacks. 
However, some classifiers, generally DNNs, can be proven robust if they produce a constant output for some set of variations of the inputs which is generally expressed as an $L_p$ ball.

DNNs have input and output layers with hidden layers between them.
Reluplex~\citep{katz2017reluplex} verifies the robustness of DNNs by searching for linear combinations of hidden layers. 
This problem is NP-complete, and thus, the search space is reduced by a simplex algorithm. 
This algorithm is based on a satisfiability modulo theories (SMT) solver that addresses Boolean satisfiability.
Exploiting the properties from the simplex, Reluplex allows inputs to temporarily violate their feasible bounds for certification, verifying the robustness of a neural network. 

Sinha et al.~\citep{sinha2017certifiable} introduced a method that is provably robust to perturbations distributed in a Wasserstein ball.
They trained a classifier with adversarial training using a distributionally robust optimization. 
Hein et al.~\citep{hein2017formal} showed formal guarantees on the robustness of classifiers using a bound on the local Lipschitz constant in the vicinity of the input. 
Their Cross-Lipschitz regularizer increased the range of attacks that can be defeated, forcing potential attackers to find better modes of attack.

Accurate bounds on worst-case losses improve the robustness. 
Raghunathan et al.~\citep{raghunathan2018certified} improved the accuracy of both the lower and upper bounds on the worst-case loss by concentrating on the upper bound.
This was performed on the basis that it is safer to minimize the upper bound than minimizing the lower bound. 
They demonstrated a novel certified approach against adversarial examples on two-layer networks.
Wong et al.~\citep{wong2018provable} presented a convex outer-bound approach called an "adversarial polytope", which is the set of all the final activation layers that are produced by applying norm-bounded perturbation to the inputs.
They used this bound for linear relaxation of the ReLU activation and optimized the worst-case loss over the region within the bound, as shown in Fig.~\ref{fig:adversarial_polytope}. 
However, this method can only be applied to small networks. 
Wong et al.~\citep{wong2018scaling} extended the scope of this method by introducing a provably robust training procedure for general networks, formulated in terms of Fenchel conjugate functions, nonlinear random projections, and model cascade techniques.

Cohen et al.~\citep{cohen2019certified} addressed the issue of certified defenses from a different perspective; they proved that classifiers that are robust against Gaussian noise are also robust against adversarial perturbations bounded by the $l_2$ norm. 
They used randomized smoothing, which had already been proven~\citep{lecuyer2019certified} to maintain robustness. 
Cohen et al.~\citep{cohen2019certified} further proved that smoothing with Gaussian noise can induce certifiable robustness against $l_2$ norm bounded perturbations. 
Because the exact evaluation of the robustness of the classifier is not possible, they showed that the method is robust against attacks with high probability using Monte Carlo algorithms.

\begin{figure}[t]
    \centering
    \includegraphics[width=1\linewidth]{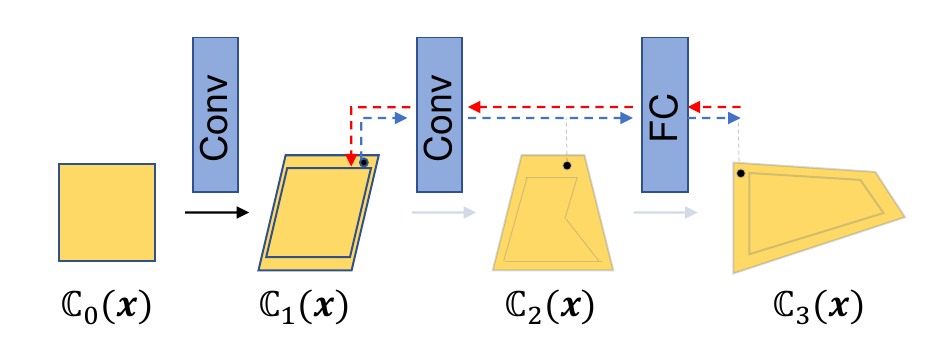}
    \caption{Illustration of layerwise adversarial training. A latent adversarial example is found in the convex region $\mathbb{C}_1(\mathbf{x})$ and propagated through the latter layers in a forward pass, which is represented by the blue lines. The red line shows the gradients during a backward pass. In the procedure, the first layer that corresponds to the former layer of convex region $\mathbb{C}_1(\mathbf{x})$ does not receive gradients~\citep{balunovic2020adversarial}.}
    \label{fig:bridging_gap}
\end{figure}

Recently, Balunovic et al.~\citep{balunovic2020adversarial} combined the adversarial training of a classifier with provable defense methods. 
A verifier aims to prove the robustness of the classifier, while an adversary attempts to garner inputs that cause errors within convex bounds, as shown in Fig.~\ref{fig:bridging_gap}. 
They utilized layerwise adversarial training and bridged the gap between adversarial training-based empirical defense methods and existing certified defense methods. 
The method resulted in state-of-the-art robust accuracy on the CIFAR-10 dataset under $2/255$ $L_\infty$ and $8/255$ $L_\infty$ perturbations.

\subsubsection{\textbf{Defense against Poisoning Attacks}}
\label{poisoning_defense}
Steinhardt et al.~\citep{steinhardt2017certified} proposes a data sanitation defense method~\citep{data_sanitation_defense} which aims to remove poisoned data points from the given dataset.
The proposed online learning algorithm provides candidate attack data instances and the worst-case test loss from any attack.

Koh et al.~\citep{koh2017understanding} used influence functions to track model predictions and identify training data points that had the most influence on a given prediction. 
Although their theory does not extend to nonconvex and nondifferentiable models, they proved that approximate influence functions can be effective against poisoning attacks.
These functions additionally allow a defender to focus on data with a high influence score. 
This method appears to be a better way of eliminating tainted examples than simply identifying data points with large training losses.

Paudice et al.~\citep{paudice2018detection} also suggested a defense mechanism to mitigate the effects of poisoning attacks based on outlier detection. 
An attacker would attempt to cause the greatest possible impact with a limited number of poisoning points. To mitigate this effect, they divided the trustworthy dataset $\mathcal{D}$ into two classes, $\mathcal{D}_{+}$ and $\mathcal{D}_{-}$, and trained a distance-based outlier detector for each class.
Each detector calculated an outlier score for each sample in the entire clean dataset. 
There are many ways to measure the outlier score, such as an SVM or local outlier factor (LOF). 
In this study, the empirical cumulative distribution function (ECDF) of training instances was used to determine a threshold for detecting outliers. 
Upon removing all the entities expected to be contaminated, the remaining data were used to retrain the learning algorithm.

Instead of following outliers, Paudice et al.~\citep{paudice2018label} decided to relabel data points that considered outliers by a label-flipping attack, which is a poisoning attack, wherein an attacker changes the label of few training points. 
They considered the points farthest from the decision boundary to be malicious and reclassified them. 
The algorithm reassigned the label of each malicious example using a k-nearest neighbor (k-NN) algorithm. 
For each sample of the training data, the closest k-NN points were first found using the Euclidean distance. 
If the number of data points with the most common label among the k-NN was equal to or greater than a given threshold, the corresponding training sample was renamed as the most common label in k-NN.

Chen et al.~\citep{chen2018detecting} looked for poisoned data by monitoring activation in the latent space of a neural network, rather than analyzing its input or output. 
Each example was analyzed by how far the degree of activation deviation from the activation-value distributions of a class majority.
Tran et al.~\citep{tran2018spectral} also defended against other variation backdoor attacks by monitoring activation values, which were analyzed using spectral signatures.
This method spotted poisoned data using the activation of a neural network, similar to the method used by Chen et al.~\citep{chen2018detecting}. 
Firstly, a singular value decomposition was applied to the covariance matrix. 
Subsequently, all the training data were compared with the first singular vector.
Poisoned examples had a high outlier score and were erased before retraining the neural network.

The defense proposed by Liu et al.~\citep{liu2018fine} was different from the mechanisms described above, which aims to detect and remove poisoned data. These authors modified the neural network itself, using a technique called fine-pruning (combination of pruning and fine-tuning).
Pruning a neural network removes neurons, including the backdoor neurons~\citep{gu2017badnets}. 
However, because other attacks are made pruning-aware, this method also suggested cleaning the neural network through fine-tuning after pruning using trusted clean data.
The resultant network was found to be robust against multiple poisoning attacks.
Wang et al.~\citep{wangneural} presented a similar method to that of Liu et al.~\citep{liu2018fine}, except they pruned filters of a neural network that were compromised, thus triggering a backdoor attack.

%% file: 1_tb-secureai.tex
\begin{spacing}{0.5}
    \setlength{\tabcolsep}{15pt}
    \ctable[
        caption = Types of attacks on secure AI and defenses techniques employed against them.,
        label   = tbl:secureai-overview,
        pos     = t,
        star
    ]{clrrr}{}{
        \toprule
        \multirow{5}[4]{*}{Attack} & \multicolumn{1}{c}{\multirow{2}[2]{*}{Evasion}} & \multicolumn{1}{l}{White-box} & & \multicolumn{1}{l}{Section \ref{whitebox_attack}} \\
        &   & \multicolumn{1}{l}{Black-box} &   & \multicolumn{1}{l}{Section \ref{blackbox_attack}}\\
        \cmidrule{2-5}      & \multicolumn{1}{c}{\multirow{3}[2]{*}{Poisoning}} & \multicolumn{2}{l}{Performance degradation} & \multicolumn{1}{l}{Section~\ref{perf_degra_attack}} \\
        &   & \multicolumn{2}{l}{Targeted poisoning} & \multicolumn{1}{l}{Section~\ref{targted_poisoning_attack}} \\
        &   & \multicolumn{2}{l}{Backdoor} & \multicolumn{1}{l}{Section~\ref{backdoor_attack}} \\
        \midrule
        \multirow{5}[4]{*}{Defense} & \multicolumn{1}{c}{\multirow{4}[2]{*}{Evasion attacks}} & \multicolumn{2}{l}{Gradient masking} & \multicolumn{1}{l}{Section~\ref{gradient_masking}} \\
        &   & \multicolumn{2}{l}{Robustness} & \multicolumn{1}{l}{Section~\ref{robustness}} \\
        &   & \multicolumn{2}{l}{Detection} &  \multicolumn{1}{l}{Section~\ref{detection}} \\
        &   & \multicolumn{2}{l}{Certification} &  \multicolumn{1}{l}{Section~\ref{certification}}\\
        \cmidrule{2-5}      & Poisoning attacks &   &   &  \multicolumn{1}{l}{Section~\ref{poisoning_defense}}\\
        \bottomrule
    } 
\end{spacing}

\begin{spacing}{1}
    \setlength{\tabcolsep}{15pt}
    \ctable[
        caption = {Attack methods against Secure AI},
        label = {tbl:secureai-attack},
        star
    ] {r|cccc} {} {
        \hline
        \toprule
        \multirow{2}{*}{}   &  \textbf{White-box} & \textbf{Black-box} & \textbf{Training phase} & \textbf{Inference phase} \\
        \textbf{Adversarial Attack Types ↓} & (Fig.~\ref{fig:evasion_attack}a)  & (Fig.~\ref{fig:evasion_attack}b) &  \\ \hline
        \textbf{Evasion}    & \checkmark   & \checkmark                &               &     \checkmark                \\
        \textbf{Poisoning}  & \checkmark   &                           &  \checkmark                                    \\
        \bottomrule
    }
\end{spacing}

%% file: 3-privateai.tex
\begin{figure*}[t]
    \centering
    \includegraphics[width=0.85\textwidth]{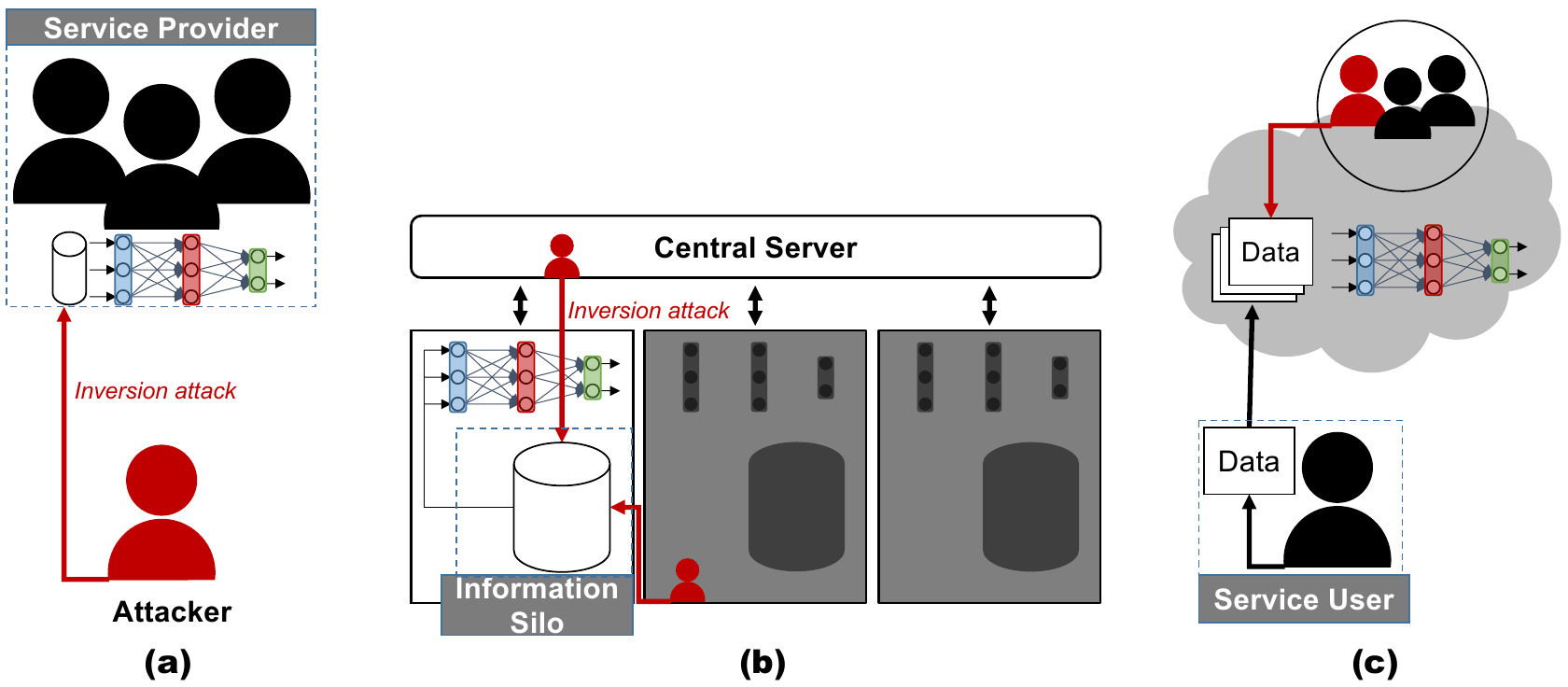}
    \caption{Privacy attack scenarios from the perspectives of the (a) service provider, (b) information silo, and (c) user.}
    \label{fig:privateai}
\end{figure*}

Deep learning algorithms, which underpin most current AI systems, are data-driven, which exposes them to privacy threats while data collection or pretrained-model distribution is performed. 
Many attempts were made to build private AI systems to maintain data privacy.
In this section, we describe the ways in which privacy can be breached in current AI systems, review defenses based on homomorphic encryption (HE), secure multi-party computation (SMC), and differential privacy (DP).

\subsection{\textbf{Scenarios for Privacy Attacks}}

\subsubsection{\textbf{Service providers}}
\label{sec:privateai-threats}
Service providers offer AI-based applications to the public.
These applications are based on pretrained DL models, and often use privacy-sensitive data to improve model performance.
A group of studies~\citep{veale2018algorithms} has suggested that not only DL models learn latent patterns from training data, but also a trained model becomes actually a repository of that data, which would effectively be exposed by granting access to a pretrained model.
In a membership inference attack~\citep{mia_shadow,mia_rnn,mia_gan,label_only,mia_white}, an attacking model tries to determine whether the given dataset was used to train the target model.
The more powerful inversion attack aims to obtain the attributes of the unknown data that were used to train the target model.
For example, Fredrikson et al.~\citep{model_inversion_confidence} reconstructed an image of a face that was used to train a target classifier using confidence scores attached to the classification.

\subsubsection{\textbf{Information Silos}}
An information silo is a data management system that is isolated from other similar systems.
A deep-learning system is usually more effective if it is trained using a large dataset.
In an AI system, information from different silos may be used to train the model without directly sharing data among the silos.
Federated learning~\citep{fed1602,fed1610,fed1902} facilitates this process by sharing gradients and model parameters; however, this makes the data vulnerable to membership and inversion attacks illustrated in Fig.~\ref{fig:privateai}.
Hitaj et al.~\citep{attack_gan} demonstrated that a federated DL approach is essentially broken in terms of privacy, because it is virtually impossible to protect the training data of honest participants from an attack in which a GAN tricks a victim into revealing sensitive data.

\subsubsection{\textbf{Users}}
\label{section:users}
Many DL-based applications run on third-party servers, because they are too large and complicated~\citep{he2016deep, huang2017densely} to run on devices such as mobile phones or smart speakers.
Users must therefore transfer sensitive data, such as voice recordings or images of faces, to the server.
Therefore, the user loses control of their data: they can neither delete it nor determine the manner in which it is used.
Similar to the recent Facebook–Cambridge Analytica data scandal~\citep{Facebook}, privacy policies may be inadequate in preventing the exploitation of user data.

\input{1_tb-privateai.tex}

\subsection{\textbf{Defenses against Attacks on Privacy}}
Several methods have been attempted to combine DL with established security techniques, including homomorphic encryption (HE), secure multi-party computation (SMC) and differential privacy (DP).
Table~\ref{tbl:privateai-defense} evaluates these techniques against the privacy threats listed in Section~\ref{sec:privateai-threats}, and in this subsection, we review their effectiveness.

\subsubsection{\textbf{Homomorphic encryption on deep learning}}
\label{section:fhe}
HE is a cryptographic scheme that enables computations on encrypted data without decryption.
An encryption scheme is homomorphic for operation $*$; without the access to the secret key, the following holds:
\begin{align}
    Enc(x_1) * Enc(x_2) = Enc(x_1 * x_2),
\end{align}
where $Enc(\cdot)$ denotes the encryption function.
HE can protect user data from third-party servers or gradients aggregated among information silos.

Gilad et al.~\citep{gilad2016cryptonets} introduced the use of encrypted data in the inference phase. 
Their CryptoNets system uses a YASHE leveled HE scheme~\citep{bos2013improved} to provide privacy-preserving inference on a pretrained CNN. 
CryptoNets demonstrated over 99\% accuracy in a classification task on handwritten digits in the MNIST dataset~\citep{lecun2010mnist}. 
However, because nonlinear activations are approximated by square functions, the extensibility of CryptoNets to large complicated models is questionable.~\citep{he2016deep,huang2017densely}.
However, Hesamifard et al.~\citep{hesamifard2017cryptodl} and Chabanne et al.~\citep{chabanne2017privacy} attempted to improve CryptoNets using higher-degree polynomial approximations of the activation functions.
Chabanne et al.~\citep{chabanne2017privacy} employed batch normalization to reduce the difference in accuracy between the original classifier and the classifier evaluated with encrypted data by approximating the activation function during inference.
This technique also permitted the design of a deeper model.
The inference on the original version of CryptoNets was slow by several hundreds of seconds; its speed was subsequently improved in later studies~\citep{chou2018faster, brutzkus2018low}.

TFHE~\citep{chillotti2016faster} is a recent HE technique that supports operations on binary data. 
TAPAS~\citep{sanyal2018tapas} and FHE-DiNN~\citep{bourse2017fast} were improvements on this scheme, implemented using with binary neural networks, which achieved improved speed and greater accuracy on the MNIST dataset, with only a single hidden layer.

\subsubsection{\textbf{Secure multi-party computation on deep learning}}
\label{section:smc}
Till date, there are two major approaches proposed to maintain privacy in DL-based systems involving multiple parties: 1) protection of user-side privacy by secure multi-party computation, and 2) secure sharing of gradients between information silos.

SMC methods are primarily based on secure two-party (2PC) techniques, which involve a user, who provides data, and a server that runs a DL system using the data. 
SecureML~\citep{mohassel2017secureml} was the first privacy-preserving method proposed in which neural networks were computed using 2PC as it requires large amounts of communication. 
In MiniONN~\citep{liu2017oblivious} a neural network is replaced by an oblivious neural network that is trained using a simplified HE scheme. 
Garbled circuits (GCs) were used to approximate nonlinear activation functions.
DeepSecure~\citep{rouhani2018deepsecure} performs encrypted inference on a neural network using Yao's GCs~\citep{yao1986generate} and suggests some other practical computing structures that are probably secure.
Gazelle~\citep{juvekar2018gazelle} performed linear operations with HE and computed the activation functions with GC. 
The authors observed that HE is the most promising method in matrix-vector multiplications, while GCs make them more suited to approximation of nonlinear functions in DNN models.
Although 2PC-based algorithms have shorter inference times than those of HE-based methods, they require communication at every operation or layer. 
Hence, they are impractical because 1) the user must be online during the whole inference phase and 2) the communication overhead increases as the number of connected users increase.

Methods to protect data privacy in the federated learning of data silos are based mainly on distributed DL algorithms~\citep{dean2012large,abadi2016tensorflow,lee2018tensorlightning}. 
The distributed selective SGD (DSSGD)~\citep{shokri2015privacy} uses collaborative DL protocols that allow different data holders to train joint DL models without sharing their training data. 
Using coordinated learning models and objectives, the participants train their local neural networks and periodically exchanged the gradients and parameters. 
As gradients and parameters are only partially shown, DSSGD resists model inversion and membership attacks. 
However, because DSSGD uses a parameter server~\citep{li2014scaling}, Aono et al.~\citep{aono2018privacy} noted that it is possible to reconstruct the data used in training from a small number of gradients. 
To preserve privacy against an honest-but-curious parameter server, the authors applied HE to the parameter and gradient exchange. 
Because the size of the encrypted data is much larger than that of plaintext, this has a trade-off with communication costs. 
Hence, Ryffel et al.~\citep{ryffel2018generic} attempted to build a privacy-preserving federated learning framework that combines MPC and DP functionality.

\subsubsection{\textbf{Differential privacy in deep learning}}
\input{3_2-2-differential_privacy}

%% file: 1_tb-privateai.tex
\begin{spacing}{\mytablespacing}
    \setlength{\tabcolsep}{15pt}
    \ctable[
        caption = {Defense methods for private AI},
        label = {tbl:privateai-defense},
        doinside = \scriptsize,
        star
    ]
    {r|ccc}
    {}
    {
        \hline
        \toprule
        \multirow{2}{*}{}  (Figure~\ref{fig:privateai}) &  \textbf{Homomorphic} & \textbf{Secure Multi-party} & \textbf{Differential} \\
        \textbf{Victims} & \textbf{Encryption}  & \textbf{Computation} & \textbf{Privacy}\\ \hline
        \textbf{Service Providers}  & & & \checkmark \\
        \textbf{Information Silos}  & \checkmark  & \checkmark & \checkmark  \\
        \textbf{Service Users}& \checkmark & \checkmark  & \\
        \midrule
        \multirow{2}{*}{\textbf{References}} & \citep{gilad2016cryptonets,hesamifard2017cryptodl,chabanne2017privacy,chou2018faster,brutzkus2018low,sanyal2018tapas,bourse2017fast} &
        \citep{mohassel2017secureml,liu2017oblivious,rouhani2018deepsecure,juvekar2018gazelle,shokri2015privacy,aono2018privacy} &
        \citep{dp_sgd,dp_sgd_rnn,dp_sgd_gan,dpgm,dp_sgd_improved,dp_fm_lr} \\
        & & & \citep{dp_fm_ae,dp_fm_cdbn,dp_pate,dp_pate_scale,dp_pate_gan}   \\
        \midrule
        \textbf{Section} & \ref{section:fhe} & \ref{section:smc} & \ref{section:dp} \\
        \bottomrule
    }
\end{spacing}

%% file: 3_2-2-differential_privacy.tex
\begin{figure}[t]
    \centering
    \includegraphics[width=1\linewidth]{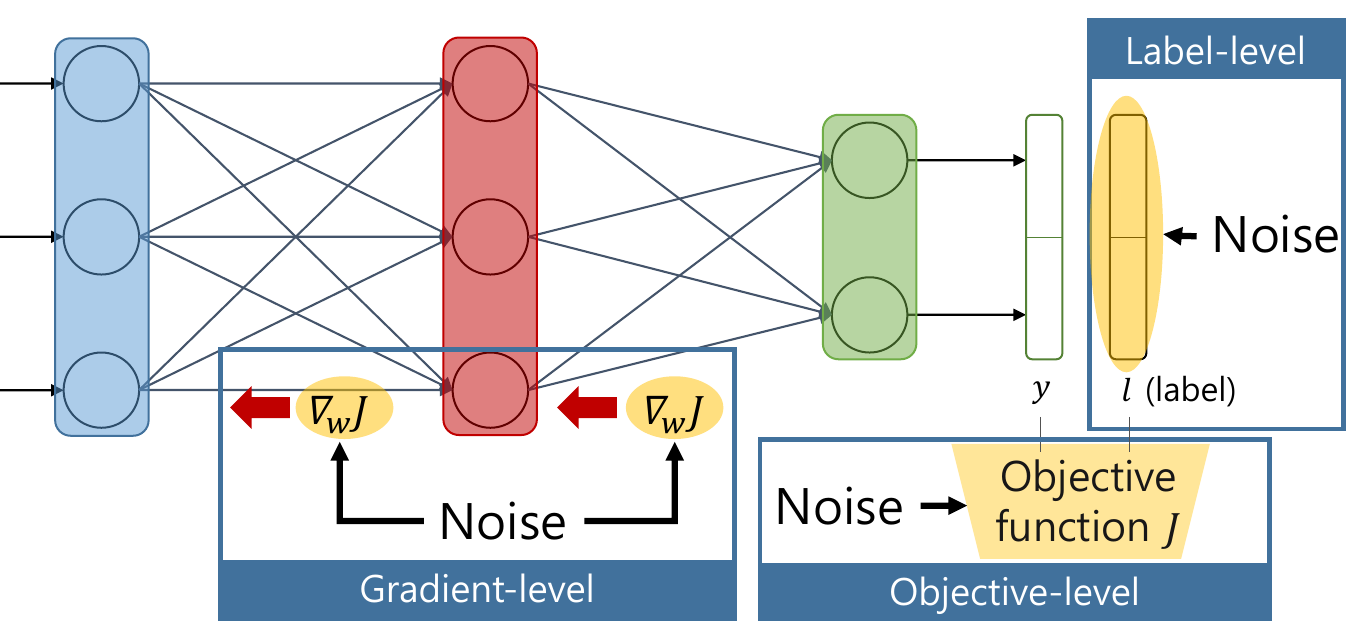}
    \caption{Overview of the differential privacy in a DL framework.}
    \label{fig:dp_dl}
\end{figure}

\label{section:dp}
Differential privacy (DP) is a state-of-the-art privacy-preserving mechanism~\citep{dp_survey} that reliably prevents an attacker from deducing private information from databases or deep learning models.
DP algorithms prevent an attacker from discovering the existence of a particular record by adding noise to the query responses, as follows:

\begin{align}
    M(\mathcal{D}) = f(\mathcal{D}) + n,
\end{align}
where $M: \mathcal{D} \rightarrow \mathbb{R}$ is a randomized mechanism that adds the noise $n$ sampled from a Laplace or Gaussian distribution~\citep{dp_survey} to the query response, $\mathcal{D}$ is the target database, and $f$ is the original deterministic query response.

$M$ provides $(\varepsilon, \delta)$-DP if all adjacent $\mathcal{D}$ and $\mathcal{D}'$ satisfy the following~\citep{dp_survey}:
\begin{align}
    \textrm{Pr}[M(\mathcal{D}) \in S] \leq \exp ({\varepsilon}) \textrm{Pr}[M(\mathcal{D}')\in S] + \delta,
    \label{eq:dp_def}
\end{align}
where $\mathcal{D}$ and $\mathcal{D}'$ are two adjacent databases, $S \subseteq \mathrm{range}(M)$ is a subset of $\mathbb{R}$, and $\varepsilon$ and $\delta$ are the privacy budget parameters that determine the level of privacy.
Smaller $\varepsilon$ and $\delta$ indicate that $M(\mathcal{D})$ and $M(\mathcal{D}')$ are more similar.

Differentially private deep learning models can be divided into three groups: gradient-level~\citep{dp_sgd, dp_sgd_rnn, dp_sgd_gan, dpgm, dp_sgd_improved}, objective-level~\citep{dp_fm_lr, dp_fm_ae, dp_fm_cdbn} and label-level~\citep{dp_pate, dp_pate_scale, dp_pate_gan} approaches, depending upon where the noise is added.
Fig.~\ref{fig:dp_dl} shows an overview of these approaches.
In the gradient-level approach, noise was inserted into the gradients of the parameters during the training phase.
In the objective-level approach, noise was used to perturb the coefficients of the original objective function.
In the label-level approach, noise was inserted into the label in the knowledge-transfer phase of the teacher--student mechanism.

Abadi et al.~\citep{dp_sgd} proposed a differential private SGD algorithm (DP-SGD) that adds noise to the gradients while updating parameters.
They introduced a \textit{moment accountant} algorithm to track the cumulative privacy loss to estimate $\varepsilon$ and $\delta$.
McMahan et al.~\citep{dp_sgd_rnn} introduced differentially private long short-term memory (LSTM)~\citep{lstm}, which provides DP for a language model.
Xie et al.~\citep{dp_sgd_gan} proposed a differentially private GAN (DPGAN) to provide DP for a differentially private generator.
The DPGAN injected noise into the gradients of the discriminator to obtain a differentially private discriminator.
The generator is trained with the discriminator, and hence becomes differentially private based on post-processing theory~\citep{dp_survey2}.
Acs et al.~\citep{dpgm} introduced a differentially private generative model consisting of a mixture of generative NNs such as restricted Boltzmann machines (RBMs)~\citep{lecun2015deep} and variational autoencoders (VAEs)~\citep{vae}.
These authors applied a differentially private $k$-means algorithm to cluster the original datasets and used DP-SGD~\citep{dp_sgd} to train each neural network.
Yu et al.~\citep{dp_sgd_improved} introduced improved DP-SGD by applying a different sampling strategy and a concentrated DP (CDP)~\citep{dp_concentrated}---a variant of DP---to provide a higher level of privacy.

The objective-level approach introduced by Chaudhuri and Monteleoni~\citep{dp_fm_lr} disturbs the original objective function by adding noise to its coefficients, making the model trained on this function differentially private.
Noise is injected into the polynomial objective function by changing the coefficients.
A non-polynomial objective function must be approximated using techniques such as the Taylor or Chebyshev expansions.
Chaudhuri and Monteleoni~\citep{dp_fm_lr} proposed a differentially private logistic regression, in which the parameters were updated to minimize perturbed objective function.
Phan et al.~\citep{dp_fm_ae, dp_fm_cdbn} applied this mechanism to autoencoders~\citep{basic_ae} and convolutional deep belief networks~\citep{basic_cdbn}.

The label-level approach injects noise into the knowledge-transfer phase of the teacher-student mechanism.
Papernot et al.~\citep{dp_pate} introduced a semi-supervised knowledge-transfer technique called private aggregation of teacher ensembles (PATE), which is a type of teacher-student mechanism whose purpose is to train a differentially private classifier (the student) based on an ensemble of non-private classifiers (the teachers), trained on disjoint datasets.
The teacher ensemble outputs noisy labels by noisy aggregation of each teacher's prediction, which the student learns.
Because the student model cannot access the training data directly and the labels that it receives are differentially private, PATE provides DP.
PATE utilizes a \textit{moment accountant} to track the privacy budget spent through the learning process.
Later, Papernot et al.~\citep{dp_pate_scale} extended PATE to operate at a large scale by introducing a new noisy aggregation mechanism, which outperformed the original PATE.
Jordon et al.~\citep{dp_pate_gan} applied the PATE to train a discriminator to build a differentially private GAN framework.
The discriminator provided DP, and the generator trained with the discriminator was also differentially private using the post-processing theory~\citep{dp_survey2}.

%% file: 4-conclusion.tex
DL has become a ubiquitous technology, and the security and privacy of DL-based systems are issues of growing importance. 
We reviewed methods of attack and defense methods in terms of model security and data privacy. 
Finally, we discuss the open problems and future research directions.

\subsection{Risks inherent in wide deployment}
It requires considerable effort to develop a new DL system; therefore, successful systems are often modified for new applications, and trained using various datasets. 
For example, U-Net~\citep{ronneberger2015u} is a CNN used for biomedical image segmentation and has a number of variants~\citep{cciccek20163d,li2017h,jo2018quantitative}.
Suppose you are a doctor who developed a new variant of U-Net trained on your patient data, the data must not be revealed to the public.
Subsequently, if you do not upload your model, the ubiquity increases the susceptibility of the CNN to the attacks reviewed in this paper. 
The availability of the U-Net and its variants can help in building substitute models for black-box attacks, and knowledge of similar models makes the model vulnerable to inversion attacks.
Hence, it is essential to conclude that the wide adoption of open-source DL systems is not compatible with high levels of security and privacy. 
There is perhaps some scope for research on the internal structuring of DL systems to permit some variations around a secure core.

\subsection{Challenges in model security}
The contributions of adversarial examples to the robustness of neural networks against misclassification attacks have recently been reexamined~\citep{ilyas2019adversarial}.
We believe that a better understanding of DL-based models is fundamental to improve both attack and defense strategies.
Evidently, interpretable AI~\citep{simonyan2013deep, basic_lrp, shrikumar2017learning} involves an analysis of the operation of a DL and the manner in which it clarifies data.
A deeper understanding of DL systems from different perspectives~\citep{ilyas2019adversarial, tsipras2018robustness} would help identify vulnerabilities to unseen attacks easily.

Moreover, as mentioned in previous studies, poisoned training data can be classified as outliers and be detected; if the poisoned data is less visible as it is highly similar to the clear data, it would be less effective. 
This provides an avenue for developing an methodology for performing a trade-off between effectiveness and detectability, which has been achieved by changing the decision boundary of the model. 
New metrics and evaluation methods are further required to secure the training phase.

In addition to designing DL-based systems as robust as current technology permits, it is essential to establish good testing and verification methods to verify it. 
For instance, there exists an automated test for DNNs used in autonomous driving vehicles~\citep{deeptest}.
Several authors~\citep{katz2017reluplex,wong2018provable,neurify,deeppoly} have presented a formal analysis of the robustness of DNNs against input perturbations. 
However, current verification methods predominantly consider only norm-ball perturbations, such as the $l_2$ ball mentioned in Section~\ref{sec:secureai}; hence, a more generalized verification is required.

\subsection{Designing privacy-preserving DNNs using HE}
The biggest problem with existing HE-based methods is that they are incompatible with DNNs.
Under current HE schemes, the number of arithmetic operations that can be performed on encrypted data is limited.
Although an operation known as \textit{bootstrapping}~\citep{bootstrapping_gentry} enables refreshing the limit, not all the HE schemes support bootstrapping, and bootstrapping generally takes a significant amount of time.
For these reasons, the existing HE-based DL systems described earlier, make their network shallow and simple to remove the necessity to perform bootstrapping; for instance, CryptoNets~\citep{gilad2016cryptonets} has only two convolutional layers and one fully connected layer, and FHE-DiNN~\citep{bourse2017fast} has only a single fully connected layer.
Therefore, designing privacy-preserving \textit{deep} neural network models using bootstrapping, which performs within a feasible time, is a fundamental future research direction.

Moreover, Classifiers that can use encrypted data are of unsuitable speeds for real-life applications. 
These systems must be implemented in parallel or distributed processing systems using graphics processing units (GPUs) or cluster processors. 
GPUs have already been used to accelerate DL training. 
Moreover, several defenses mentioned in Section~\ref{section:fhe} require the user to encrypt their data before transmission and decrypt the results. 
Hence, methods for verifying the practicability on-device encryption and decryption and improving efficiency should be devised.